\documentclass[%
 reprint,
superscriptaddress,
showpacs,preprintnumbers,
 amsmath,amssymb,
 aps,
prb,
]{revtex4-1}

\usepackage{graphicx}
\usepackage{dcolumn}
\usepackage{bm}

\begin{document}

\preprint{APS/123-QED}

\title{Tunneling spectroscopy of phosphorus impurity atom on Ge(111)-(2x1)
surface. \\ {\it Ab initio} study. }

\author{S.~V.~Savinov}
\email{SavinovSV@mail.ru}
\author{A.~.I~.Oreshkin}
\affiliation{
 Faculty of Physics, Moscow State University, 119991 Moscow, Russia\\
}

\author{S.~I.~Oreshkin}
\affiliation{
 Sternberg Astronomical Inst., Moscow State University, 119991 Moscow, Russia\\
}

\date{\today}

\begin{abstract}
We have performed the numerical modeling of Ge(111)-(2x1) surface
electronic properties in vicinity of P donor impurity atom located near the
surface. We have found a notable increase of surface $LDOS$ around surface
dopant near the bottom of empty surface states band $\pi^*$, which we called
split state due to its limited spatial extent and energetic position inside the
band gap. We show, that despite of well established bulk donor impurity energy
level position at the very bottom of conduction band, surface donor impurity on
Ge(111)-(2x1) surface might produce energy level below Fermi energy, depending
on impurity atom local environment. It was demonstrated, that impurity, located
in subsurface atomic layers, is visible in STM experiment on Ge(111)-(2x1)
surface. The quasi-1D character of impurity image, observed in STM experiments,
is confirmed by our computer simulations with a note that a few $\pi$-bonded
dimer rows may be affected by the presence of impurity atom.

%
%
\end{abstract}


\maketitle


\section{Introduction}

At present it is a common place that Ge(111)-(2x1) surface consists of
$\pi$-bonded zigzag chains. This was confirmed many times by different means,
see~\cite{Ge_bands, Ge_bands1, Bussetti2008} and references
therein. Surprisingly, just a few publications are devoted to investigations of
impurity atoms on (111) surface of elemental semiconductors~\cite{Si:P, Si:P_1,
Si:P_2, Si:P_buried, JETP_LETT2005, Si:B, Si:Bi, Si:Bi_1}. And as one can see
the interest to Si(111)-(2x1) surface is renewed. But not to Ge(111)-(2x1)
surface. This is unexplainable, because Ge is the main candidate for technology,
allowing to overcome scaling limits of Si-based MOSFETs~\cite{Ge_spin}. The
knowledge of local properties of Ge, especially caused by impurity atoms, is of
vital importance. Besides, the surface and interface properties of Ge(111) have
great significance for Ge spintronics applications. It is known Ge has some
advantages above Si~\cite{Spin}.

The STM method till now is the only physical method achieving atomic
resolution in real space. But experimenters often suffers from the lack of
some reference points provided by the theory. For example, the reliable STM
image interpretation still remains the challenging task. There is no general
approach which takes into account all kind of physical processes responsible for
STM image formation. Below we report on surface electronic structure
investigation performed by {\it ab initio} computer simulations in the
density functional framework, which is the first order estimation for STM/STS
images, and could serve as a basis for further model improvements.

We restrict the present investigation to the case of left (negative) only
surface buckling (Fig.~\ref{Bands}(I)) as at present this matter is still
controversial and is the subject of intensive investigations~\cite{Buckling,
Buckling1, Buckling2}. Our research is in some sense similar to reported
in~\cite{Si:P} for Si(111)-(2x1) surface. But we take into account
all possible impurity positions in two surface bi-layers of Ge(111)-(2x1)
reconstruction, and besides, our analysis is not aimed on pure STM image
simulation, rather on comprehensive analyses of local density of states.

\section{Methods}

We have performed our DFT calculations in LDA approximation as implemented in
SIESTA~\cite{SIESTA} package. The use of strictly localized numerical atomic
orbitals is necessary to be able to finish the modeling of large surface cell in
reasonable time. The surface Ge(111)-(2x1) super-cell consists of 7x21 cells of
elementary 2x1 reconstruction, each 8 Ge atomic layers thick (total 2646 atoms).
Vacuum gap is chosen rather big - about 20~\AA. Ge dangling bonds at the slab
bottom surface are terminated with H atoms to prevent surface states formation.
The geometry of the structure was fully relaxed, until atomic forces have became
less then 0.003~eV/\AA. More details about calculations can be found
elsewhere~\cite{JETP_LETT2011}.

As we have reported earlier, the atomic structure of Ge(111)-(2x1) surface is
strongly disturbed in vicinity of surface defects~\cite{JETP_LETT2011,
JETP_LETT2012}. A few
$\pi$-bonded rows around the defect are affected. That is why the
geometry relaxation has been performed with the large super-cell to keep the
internally periodic for DFT images of impurity well separated. Although it is
still the open question, if the defect's images separation is sufficiently
large.

At the last step of simulation the spatial distribution of Khon-Sham
wave-functions and corresponding scalar field of surface electronic density of
states $LDOS(x, y, eV)$ were calculated. Because of strictly localized atomic
orbitals, used in SIESTA, the special procedure of wave-functions extrapolation
into the vacuum has to be used (it is also implemented in SIESTA package).

\section{Results}
\subsection{Geometry and ground state properties}

In STM method the LDOS is measured {\it above} the surface, The tails of
wave-functions are actually making the image. Thus in DFT calculations we
are interested in the following quantity:

$$LDOS(x,y, eV) \sim \sum |\Psi(x,y)|^{2} \, \tilde{ \delta}(E -
E_i)|_{z=Const},$$

\noindent where $\Psi$ are Khon-Sham eigenfunctions, $\tilde{ \delta}$ is finite
width smearing function, $E_{i}$ are Khon-Sham eigenvalues, and summing is
evaluated at certain plane ($z=Const$), located a few angstroms above the
surface. Here the broadening is the essential part of calculations, as we know
from our experience that tunneling broadening in STM experiments on
semiconductors typically amounts about 100~meV. Broadening provides the degree
of $LDOS$ smoothing, necessary to resemble experimental tunneling spectra.

\begin{figure}[!h]
\leavevmode
\centering{
\includegraphics[width=60mm]{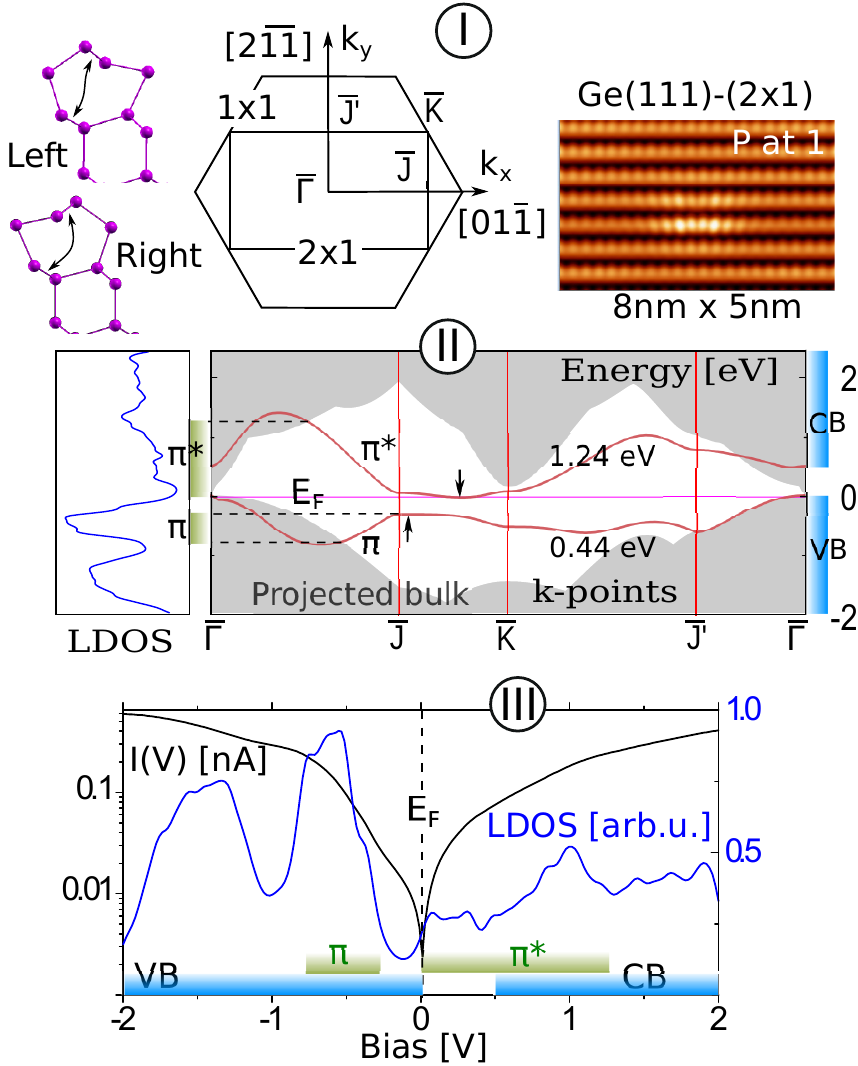}}
\caption[]{\label{Bands}
{\bf (I)} Sketches of left and right isomers of Ge(111)-(2x1) surface. Two
bonds, appropriate for definition, are marked by arrows. IBZ for Ge(111) 1x1 and
2x1 surfaces together with special points and relevant directions denoted. Model
surface slab with impurity atom positioned at site 1 (see below) correctly
oriented with respect to crystallographic directions.
{\bf (II)} Surface band diagram for Ge(111)-(2x1) reconstruction. Empty $\pi^*$
and filled $\pi$ surface states bands can be seen in projected band gap. Their
energetic position is schematically shown at ordinate axis by rectangles. To
illustrate the relation between surface band structure and $LDOS(eV)$ curve
(III), the latter is shown at the left pane. $\pi^*$ bottom and $\pi$ top are
marked by arrows. Bulk bands energetic position is shown at the right side.
{\bf (III)} The $LDOS(eV)$ curve and I(V) curve on logarithmic scale, averaged
above the whole 8nm~x~5nm surface slab shown on (I).}
\end{figure}

At the first stage of DFT calculations the equilibrium geometry has to be
established in the unit cell of Ge(111)-(2x1) reconstruction. Afterwards the
unit cell is enlarged to the desired extend, the defect is introduced and the
structure is relaxed again. The final step is  $LDOS(x, y, eV)$ calculation.
The results are sketched in Fig~\ref{Bands}.

The 1x1 and 2x1 surface irreducible Brillouin zones (IBZ) together with special
points and directions are shown on pane~{ (I)}. The right half of pane (I)
contains the $LDOS$ image correctly oriented with respect to special directions.
The image corresponds to the whole 7x21 super-cell of 2x1 surface unit cells of
Ge(111) surface, with P impurity atom placed at the position 1 (see below). We
have to note that all figures are intensively cross-referenced, and the meaning
of some notations on one figure can be better understood with another figure in
mind.

Pane Fig.~\ref{Bands}(II) illustrates electronic structure of clean
Ge(111)-(2x1) surface. Two surface states (SS) bands, empty $\pi^*$ and filled
$\pi$, can be seen in projected band gap. The widths of SS bands, derived from
Fig.~\ref{Bands}(II), equal to $\Delta \pi^*$~=~1.24~eV and $\Delta
\pi$~=~0.44~eV respectively. The surface band gap $\Delta E_{SBG}$ is
about~0.3~eV. As it should be expected, LDA approximation gives the band gap
value which is much smaller than the experimental one. SS bands, as well as
bulk bands, are also shown in Fig.~\ref{Bands}(II) by small rectangles next to
ordinate axis. This representation will be used on majority of figures below.

The surface band structure is presented for the case of Ge(111)-(2x1) surface
with negative buckling. Our calculations predict that negative surface isomer
is energetically (by almost 11~meV per (2x1) unit cell) more favorable and, as
we stated above, we will restrict present analyses to negative buckling only.

The valence band (VB) top and empty SS bottom coincide with Fermi energy in
Fig.~\ref{Bands}(II). The band gap is completely covered by empty SS $\pi^*$
band. Specific for such  band diagram was discussed in~\cite{JETP_LETT2005}.

Jumping ahead, on pane Fig.~\ref{Bands}(III) we display $LDOS(eV)$ curve.
The calculated I(V) dependence on a logarithmic scale is shown alongside.
Surface $LDOS(eV)$ curve is the result of averaging over the whole
8~nm~x~5~nm area on Fig~\ref{Bands}(I). The gap right below the Fermi level is
clearly observed on $LDOS(eV)$ graph. It does not corresponds to bulk band gap.
The closest resemblance can be found with the surface band gap.

Rectangles at absсissa axis illustrate different band's energy position. It is
necessary to state, that everywhere below, the bottom of CB is schematically
shown on the figures for the case of Ge(111)-(2x1) surface at room temperature.
In this case the optical band gap is about 0.5~eV~\cite{Ge_BG}. The DFT band gap
in LDA approximation is non-physically small, less then 100~meV. To give clear
impression on relation between surface band structure and $LDOS$, the latter is
also depicted at the left side of surface band diagram Fig.~\ref{Bands}(II).

\subsection{$LDOS$ scalar field representation}

In further exposition we will focus mostly on $LDOS$ properties and before we go
to the main results we have to clarify the physical meaning of our data
representation for $LDOS$.

\begin{figure}[!h]
\leavevmode
\centering{
\includegraphics[width=80mm]{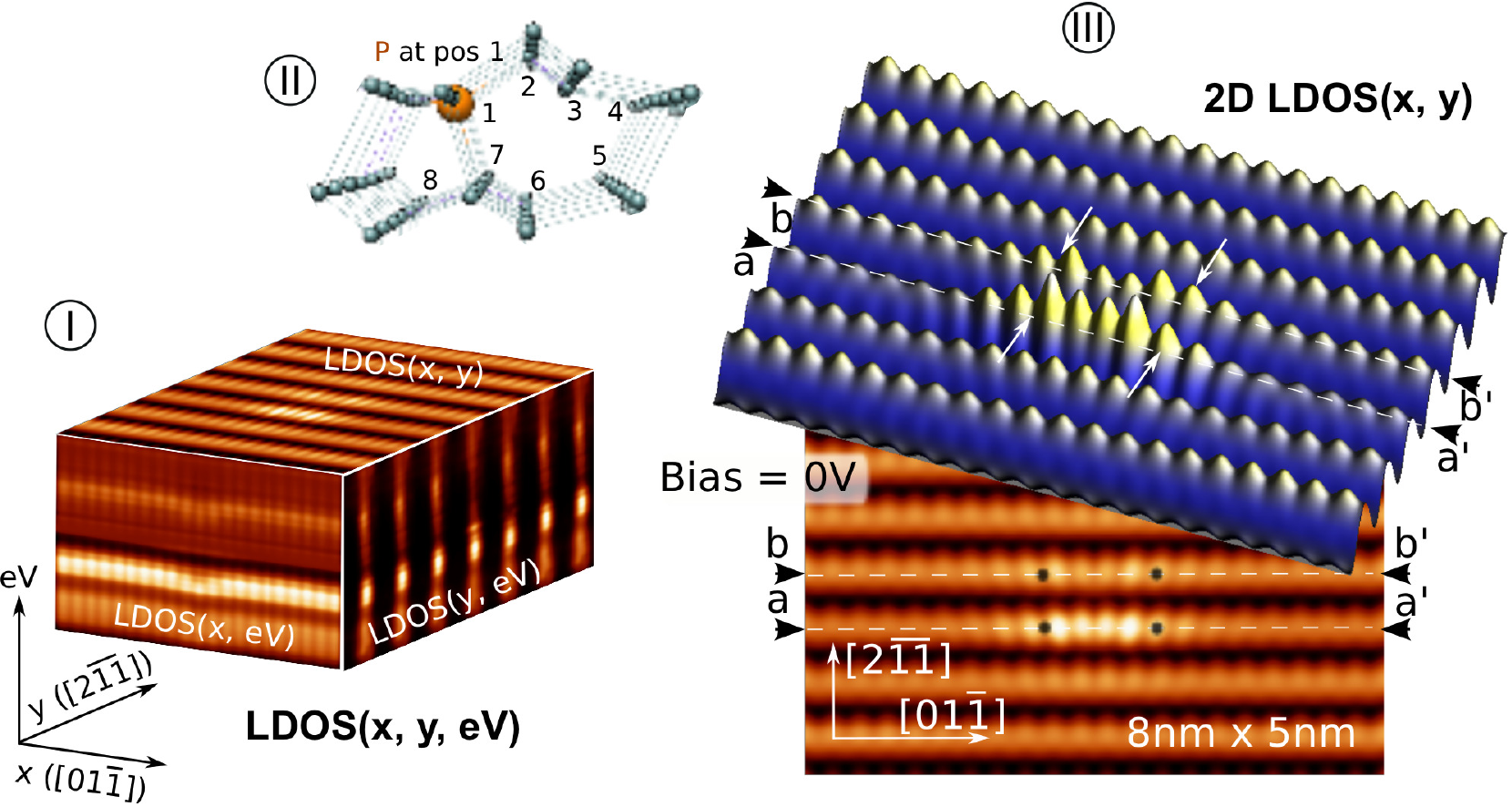}}
\caption[]{\label{Overview}
{\bf (I)} Sketch of $LDOS(x, y, eV)$ scalar field. Relevant directions are
shown.
{\bf (II)} Labels to identify donor atom position in two surface bi-layers of
Ge(111)-(2x1) reconstruction. P atom is shown in position 1.
{\bf (III)} Quasi-3D and 2D representations of $LDOS(x, y)|_{eV=0}$
cross-section of $LDOS(x, y, eV)$ scalar field at zero bias. Specific points
and directions, referred to below, are denoted.}
\end{figure}

Below we are speaking about cross-sectioning of $LDOS(x, y, eV)$ scalar field
(Fig.~\ref{Overview}(I)). The x and y directions correspond to
$[01\overline{1}]$ and $[2\overline{11}]$ crystallographic directions. The two
most relevant quantities are cross-section of scalar field $LDOS(x, y, eV)$
along $(x, y)$ and $(x, eV)$ planes - $LDOS(x, y)$ and $LDOS(x, eV)$
respectively.

The $LDOS$ is build up for different impurity atom positions in two subsurface
bi-layers of Ge(111)-(2x1) reconstruction. The definition for donor atom
position's notation is presented in Fig.~\ref{Overview}(II). Worth to mention,
that impurity atom does not occupy exactly the same lattice site as host atom
it substitutes.

In Fig.~\ref{Overview}(III) we show the cross-sections $LDOS(x, y)|_{eV=0}$ at
fixed bias voltage for the case of P donor atom located at position 1 in surface
bi-layer. These images roughly correspond to experimental STM images, as at
small bias voltage there are not too many sharp $LDOS$ features, contributing to
the image. It can be seen from Fig.~\ref{Overview}, that {\it two} $\pi$-bonded
rows of surface reconstruction are influenced by impurity. In each row a
protrusion can be observed. The spatial extent of impurity induced feature along
the direction of $\pi$-bonded dimer row ($\rm{[01\overline{1}]}$ direction) is
at least 40 \AA. Note two distinguishable maxima on the protrusion. We will come
back to this fact later. Arrows and dots on the image, as well as (a-a') and
(b-b') lines, mark spatial points and directions, referred to on figures below.

\begin{figure}[!h]
\leavevmode
\centering{
\includegraphics[width=80mm]{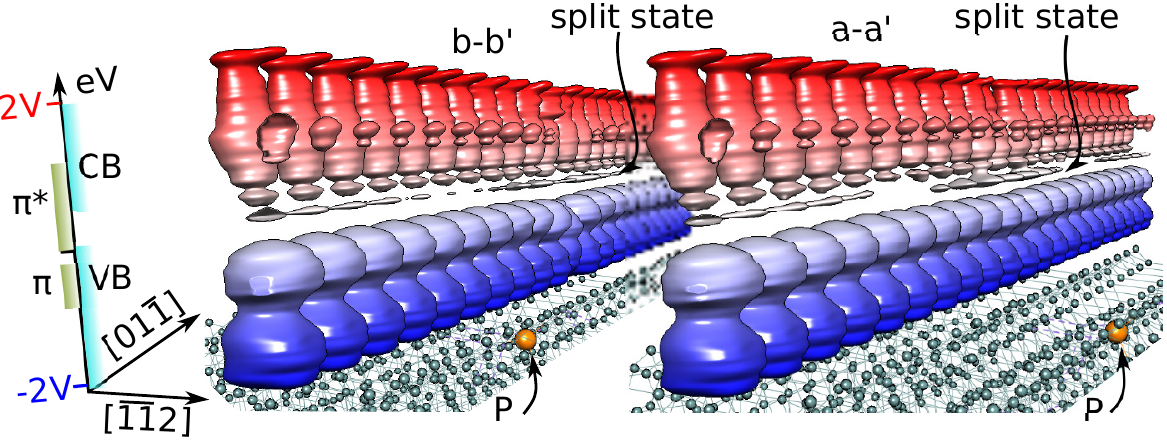}}
\caption[]{\label{LDOS_surface} $LDOS(x, y, eV)$ field above rows a-a' and b-b'
(see Fig.~\ref{Overview}) shown by surfaces of constant value. Coloring
corresponds to applied bias voltage. Split state in the band gap can be clearly
observed above both dimer rows. Note the changes of structure at the right side
of CB features and strong hybridization of atomic orbitals around impurity
atom.}
\end{figure}

We have found notable increase of $LDOS$ in vicinity of impurity atom at the
bottom of empty SS band $\pi^*$. We will refer to it as to split state.
Fig.~\ref{LDOS_surface} proves the validity of this terminology. In the figure
the $LDOS(x, y, eV)$ field is shown by surfaces of equal value, colored by
applied tunneling bias voltage. The $LDOS$ is drawn for two $\pi$-bonded rows
denoted as a-a' and b-b' in Fig.~\ref{Overview}. The position of donor atom
is depicted in the figure. To prevent the confusion caused by quasi-3D
picture, we clarify that the impurity is located at the left side for
$\pi$-bonded row b-b', and at the right side for $\pi$-bonded row a-a'.

Let us point out some important facts about this $LDOS$ representation. First of
all, high values of $LDOS$ are confined within areas bounded by surface of
constant value. They are perfectly localized above up $\pi$-bonded rows. This is
the reason, why only every second dimer row is imaged by STM. In between
$\pi$-bonded rows $LDOS$ is relatively low. Beside this, all round shaped
vertical structures in $\pi$-bonded row are located above up-atom. Down-atom can
be found in between them. Such spatial structure of $LDOS$ is a consequence
of collective $\pi$-bonds formation. STM can only image the up-row dimers, and
therefore, only up-rows. Basically, with used approach $\pi$-bonds can also
be directly visualized~\cite{JETP_LETT2011}.

The hybridization of atomic orbitals is clearly visible from
Fig.~\ref{LDOS_surface}. It is very strong in close proximity to surface defect.
This causes the appearance of specific feature near the bottom of empty SS band
(marked by arrows in Fig.~\ref{Overview}). The limited spatial and energy
extents of this feature as well as its position in the band gap implies that
this indeed is the split state.

We are working with microscopic picture, on the level of individual atoms. That
is why we are able to see the connection between atomic orbital's hybridization
and macroscopic band structure. Surface states itself appear due to atomic
arrangement of the surface. Split state appear due to changes of this
arrangement around the defect. They both have the same root - hybridization
of atomic orbitals. One problem exists. It is really difficult to define
the energetic position reference level, is it Fermi energy, or the SS $\pi^*$
band bottom? To be accurate we will refer to Fermi level, i.e. split state is
located at the Fermi level, and not near $\pi*$ bottom.

\begin{figure}[!h]
\leavevmode
\centering{
\includegraphics[width=70mm]{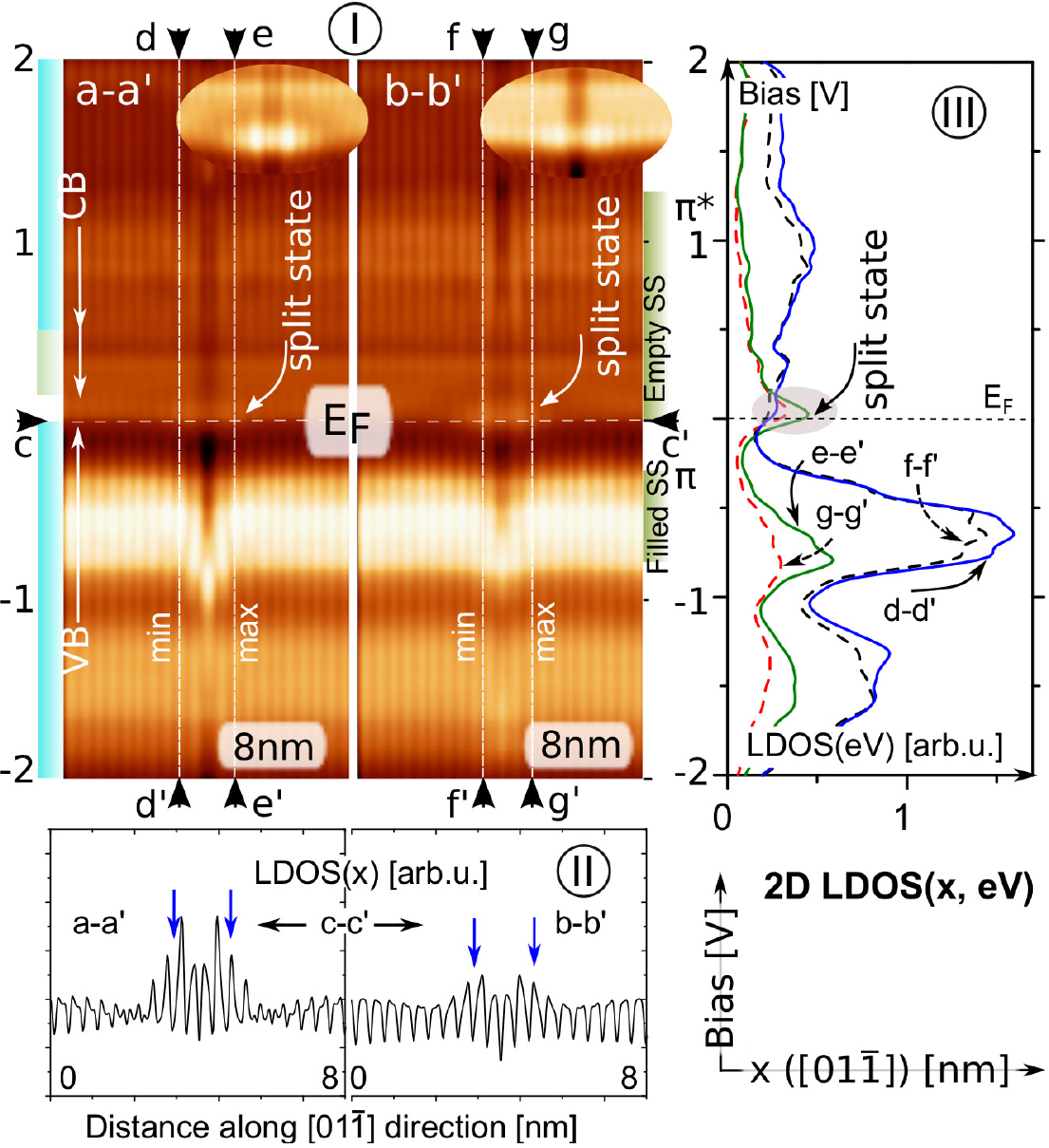}}
\caption[]{\label{LDOS_map} {\bf (I)} 2D $LDOS(x, eV)$ distributions
taken along a-a' and b-b' {\it planes} in Fig.~\ref{Bands}. Impurity atom is at
1 site. Two rectangles at the left side correspond to optical and DFT
conduction band positions. {\bf (II) } Profiles of a-a' and b-b' panes along
c-c' line, which are {\it essentially the same} as profiles of images in
Fig.~\ref{Bands} along a-a' and b-b' lines. {\bf (III)} Profiles of a-a' and
b-b' images along d-d', e-e', f-f' and g-g' lines, which are $LDOS(eV)$
dependencies at points, shown in Fig.~\ref{Bands} by black dots and white
arrows. For details see the text.\\
Axis directions and images size are indicated on the figure.}
\end{figure}

Important for our result's understanding cross-sections
$LDOS(x,eV)|_{a-a'(b-b')}$ at fixed $y$ coordinate  are shown in
Fig.~\ref{LDOS_map}(I). They are taken along (a-a') and (b-b') {\it planes}
in Fig.~\ref{Overview}, i.e. along $\pi$-bonded rows of Ge(111)-(2x1)
reconstruction. Areas near the Fermi level, where split state resides, are
zoomed in on the insets. The positions of Fermi level $\rm{E_F}$,
conduction band (CB) bottom, valence band (VB) top and empty $\pi^*$ and
filled $\pi$ surface states bands are indicated in Fig.~\ref{LDOS_surface}.
According to our DFT calculations the top of VB almost coincide with the bottom
of empty surface states band $\pi^*$ and Fermi level Fig.~\ref{Bands}
and~\cite{JETP_LETT2011}. Even more, the $\pi^*$ band can be partially filled at
very high doping ratio~\cite{Nicolls}. Our calculations also predict for
Ge(111)-(2x1) surface with negative buckling the position of empty SS $\pi^*$
band bottom about tens meV {\it below} the Fermi level.

The proportions of $LDOS(x,eV)$ images are chosen on purpose in a way that is
convenient for experimenters. Typically the number of point along spatial
direction is less than the number of bias voltage points and tunneling spectra
image is elongated in vertical direction. Though the DFT band gap in LDA
approximation is non-physically small for the sake of completeness it is also
shown in Fig.~\ref{LDOS_map}(I).

The distribution $LDOS(x, eV)$ is reach of features. Its cross-section along $x$
coordinate gives the $LDOS(x)$ profile {\it exactly} in the same way as the
cross-sectioning of $LDOS(x, y)$ (Fig.~\ref{Bands}) along $x$ coordinate. The
cross-sections of a-a' and b-b' panes along c-c' line are shown on the panel
(II). One can see two distinct maxima on the profiles. And what is really
important is that the spatial extent of perturbation is obviously about 80~\AA.
The 40~\AA~ estimation from Fig.~\ref{Overview} suffers from unsufficient
contrast of $LDOS(x, y)|_{eV=0}$ image.

Cross-section of $LDOS(x, eV)$ along $eV$ coordinate (d-d', e-e', f-f' and g-g'
lines Fig.~\ref{LDOS_map}(I)) corresponds to point spectroscopy $LDOS(eV)$
dependencies (Fig.~\ref{LDOS_map}(III)) at points of $LDOS(x)$ profile, marked
by vertical arrows in Fig.~\ref{LDOS_map}(II). These are points shown in
Fig.~\ref{Bands} by dots and arrows.

Curves d-d' and f-f' are taken in between dimers in $\pi$-bonded row, while
curves e-e' and g-g' are taken on top of dimers (Fig.~\ref{LDOS_map}(II),
Fig.~\ref{Bands}). For the whole range of bias voltage the values of $LDOS$
collected in between dimers are higher than that on top of dimers, except for
narrow interval in vicinity of Fermi energy, where resides the split state.
This split state contributes to the increase of $LDOS$ on top of dimers in
$\pi$-bonded row. Thus, the protrusion, consisting of few dimers appear on
$LDOS(x, y)$ (as well as on STM) image. It follows from Fig.~\ref{LDOS_map}(III)
that the contrast of protrusion is higher on (a-a') plane than on (b-b') plane,
and this indeed can be observed in Fig.~\ref{Overview}.

\begin{figure}[!h]
\leavevmode
\centering{
\includegraphics[width=50mm]{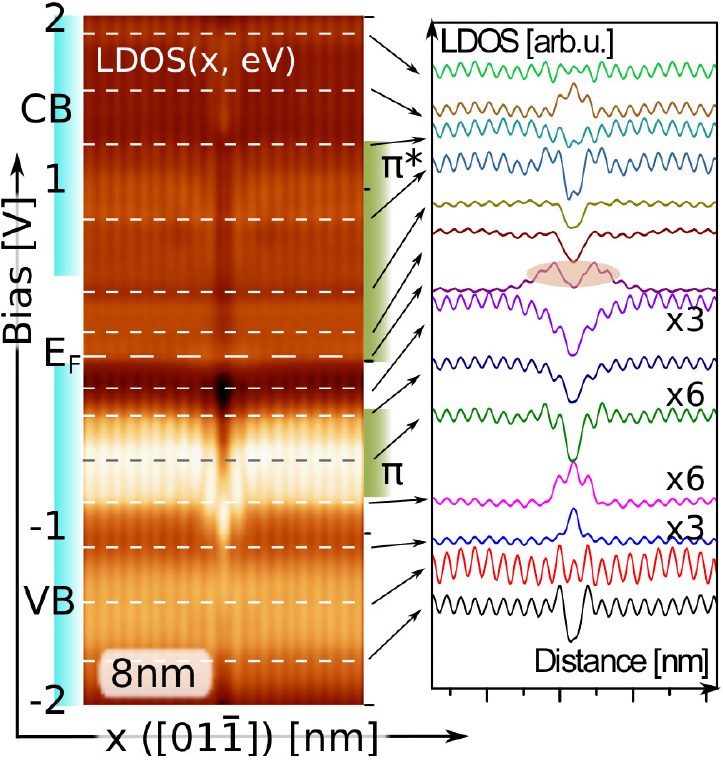}}
\caption[]{\label{LDOS_sect} $LDOS(x, eV)$ map along b-b' dimer row in
vicinity of P atom, located at positions 1 on Ge(111)-(2x1) surface and its
cross-sections along denoted lines. Note that impurity can appear on $LDOS$
image as protrusion, depression, protrusion superimposed on depression etc.}
\end{figure}

To illustrate the usefulness of $LDOS(x, eV)$ map the set of cross-sections
along spatial coordinate is depicted in Fig~\ref{LDOS_sect} for P donor atom
located at position 1. Profiles are slightly low pass filtered to stress the
long range features, so they look a bit different comparing to
Fig.~\ref{LDOS_map}(II). When tunneling bias changes, the $LDOS(x)$ profile
also changes revealing depdf and protrusions of different shape. The profile,
corresponding to split state energy (and to the presence of protrusion on STM
image) is marked by ellipse. One can easily see that the amplitude of protrusion
at Fermi level is much less than the amplitude of features at other bias voltage
(see also Fig.~\ref{LDOS_map}(III)).

Also note, that impurity's $LDOS$ image might have elongated hillock like shape
at positive bias (empty states). It means the protrusion on the $LDOS$ image is
not caused by charge density effects (like charge screening).

To the best of our knowledge this fact was never clearly stated. In other words,
STM image of Ge(111)-(2x1) surface~\cite{JETP_LETT2005} (as well as
Si(111)-(2x1) surface~\cite{Si:P, Si:P_1, Si:P_2}) around surface defect is
dominated by the split state in vicinity of Fermi level, although the amplitude
of the effect is relatively small.

\begin{figure}[!h]
\leavevmode
\centering{
\includegraphics[width=60mm]{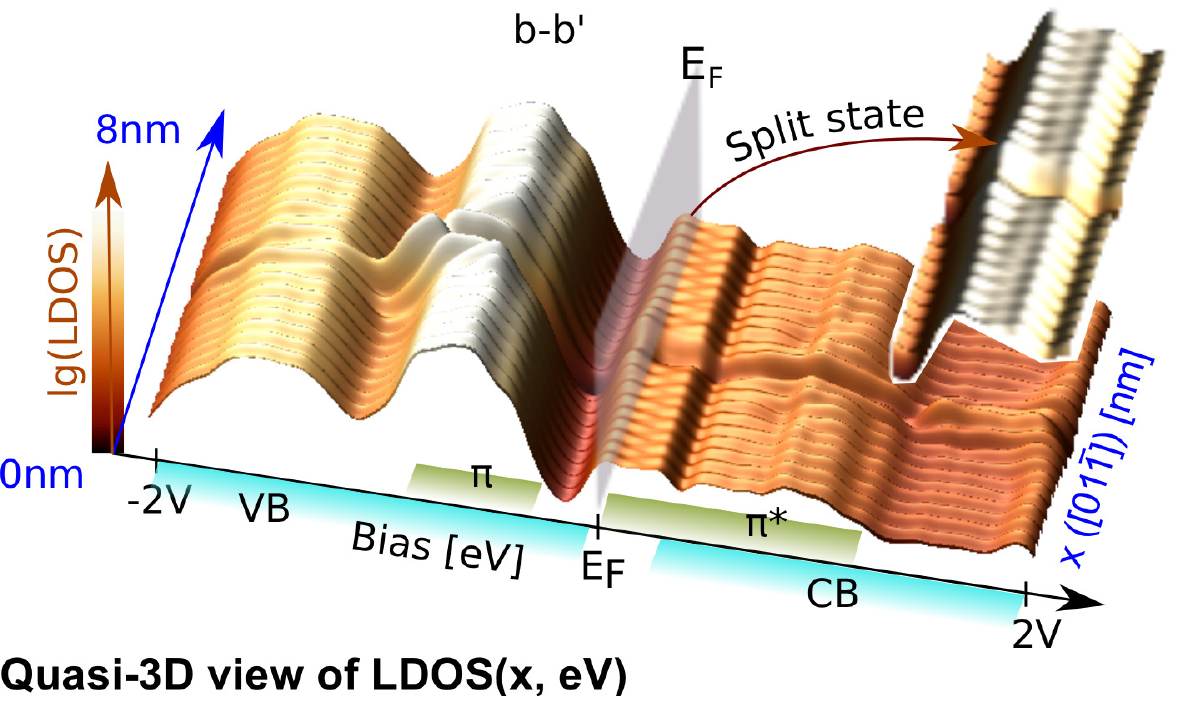}}
\caption[]{\label{LDOS_3D} Quasi-3D representation of $LDOS(x, eV)$ along {b-b'}
$\pi$-bonded row of Fig.~\ref{Bands}. Area, containing split state is zoomed in
to give clear impression about its spatial structure. Fermi level is shown as
semitransparent plane. $LDOS$ values are given on a logarithmic scale to
increase height contrast. The values of LDOS are coded both by height and by
color with lightning.}
\end{figure}

To give even more insight into the power of $LDOS(x, eV)$ data representation it
is drawn in Fig.~\ref{LDOS_3D} as quasi-3D surface. Height is given on
logarithmic scale to increase the image height contrast. The value of LDOS is
coded both by height and by color with lightning. The spatial and energetic
positions of specific features of tunneling spectrum can be easily deduced from
the figure. The split state (zoomed in on the inset) is located at Fermi level.
It has cigar like spatial shape which directly reflects in the shape of
protrusion on $LDOS(x, y)$ image.

It is also obvious from Fig.~\ref{LDOS_3D}, that split state really fills in the
whole width of $LDOS(x, eV)$ spectra. Thus we can not completely exclude the
possibility of impurity induced electronic features overlap between neighboring
super-cells of calculation. This overlap can introduce some difficult
to estimate errors in quantum mechanical forces evaluation. This is the main
reason, why we have increased the size of geometry relaxation surface cell to
the upper available to us limit.

Let us now finish the overview of data representation. Taking stated above into
account one can conclude that, given $LDOS(x, eV)$ it is readily possible to
estimate the outlook of point spectroscopy curves as well as the shape of
spatial profiles. That is why the results of electronic properties calculations
for all 8 possible positions of P impurity atom in two surface bi-layers
(Fig.~\ref{Overview}) are presented in Fig.~\ref{LDOS_maps} as $LDOS(x, eV)$
maps.

\subsection{Surface $LDOS$ around P donor impurity}

Let us point out the most important features of calculated images.

As we have discussed earlier, the empty surface states band $\pi^*$ is governing
STM image formation for Ge(111)-(2x1) surface~\cite{JETP_LETT2005,
JETP_LETT2007} in the band gap region. Now the same concerns the split state.
The STM image is dominated by
split state in vicinity of Fermi energy.

\begin{figure}[!h]
\leavevmode
\centering{
\includegraphics[width=70mm]{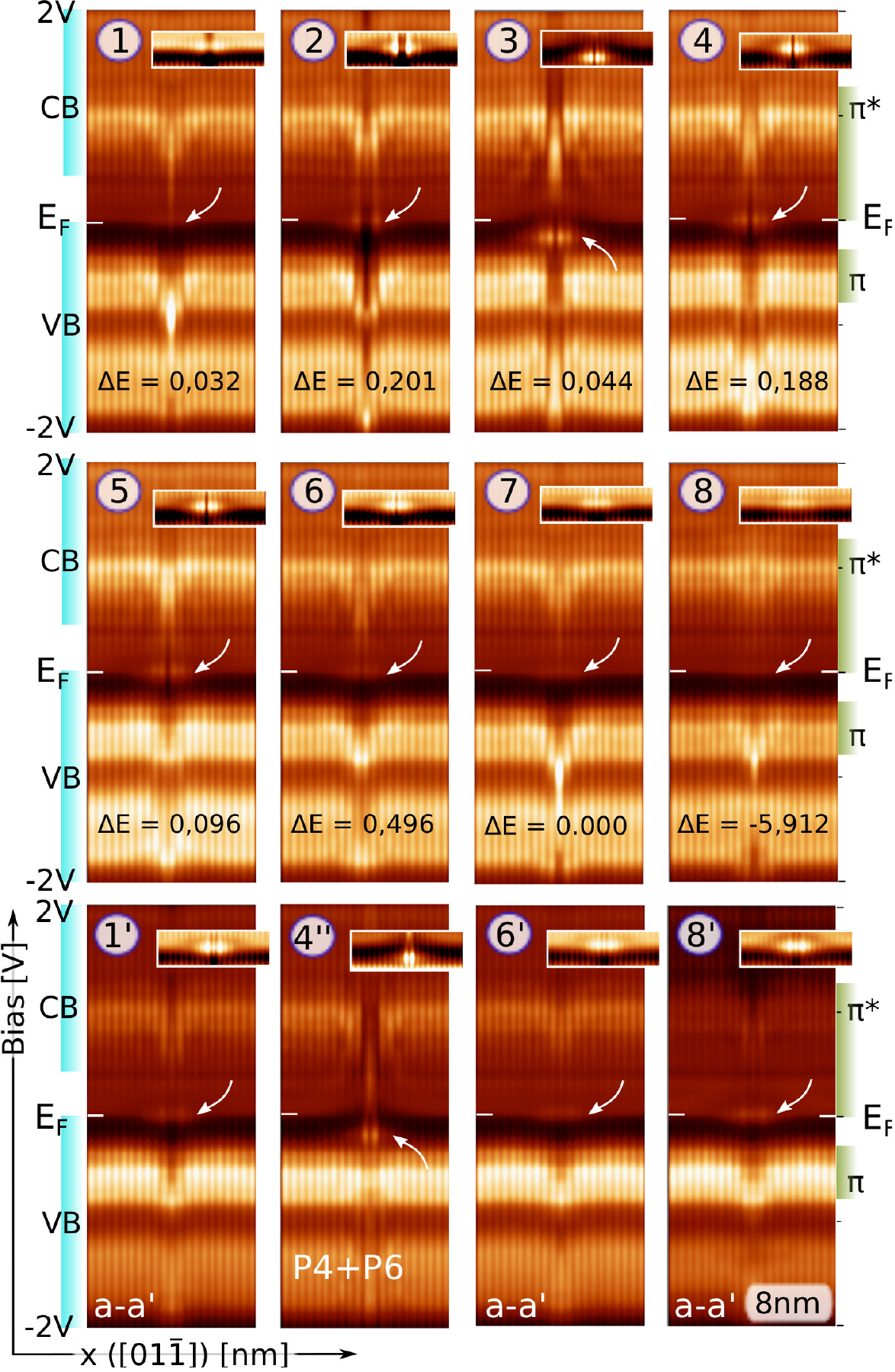}}
\caption[]{\label{LDOS_maps} {\bf (1-8)} $LDOS(x, eV)$ maps along b-b' dimer row
in vicinity of P atom, located at different positions in subsurface layers of
Ge(111)-(2x1) surface. Numbers denote atom position (see Fig.~\ref{Overview}).
Energy difference in eV relative to position 7 as well as split state position
are indicated in the figure. Area of split state is zoomed in on every pane.\\
{\bf (1', 6', 8')} $LDOS(x, eV)$ maps along a-a' dimer row for the case when two
$\pi$-bonded rows are disturbed by impurity atom. {\bf (4'')} $LDOS(x, eV)$ map
for the case of two donor atoms located at positions 4 and 6. Note split state
located {\it below} Fermi level. Axis directions and images size are indicated
near the bottom of the figure.}
\end{figure}

The noticeable influence of surface states inside CB and VB bands can be
inferred from Fig.~\ref{LDOS_maps}. There are $LDOS$ peculiarities near the top
of empty surface states band $\pi^*$ and at the edges of filled surface states
band $\pi$. They are imaged as horizontal bright stripes.

For all P doping atom positions except position 3, the split state is located at
the Fermi level. When P impurity is placed at position 3, the split state can be
observed below Fermi energy Fig.~\ref{LDOS_maps}(3). Position 2 is somewhat
specific. In this case the impurity atom is directly breaking $\pi$-bonded
chain, and this strongly influences $LDOS(x, eV)$ (Fig.~\ref{LDOS_maps}(2)) - at
almost all possible bias voltage values the impurity LDOS image has two well
pronounced peaks.

As with other semiconductors, individual impurity is visible in STM experiment 
when it is located {\it below} the Ge(111)-(2x1) surface
(Fig~\ref{LDOS_maps}(5-8)). To the best of our knowledge we report this for the
first time. Albeit it should be rather obvious from simple speculations. The
crystal lattice is disturbed noticeable far from atomic size
defect~\cite{JETP_LETT2012}, and this disturb the perfectness of collective
$\pi$-bonding  in a few dimer rows. There is also another possibility. The P is
shallow impurity in Ge, its ionization energy is 13 meV. Therefore it
localization radii should be large, in 50~\AA~ range. The $LDOS$, observed by
STM for Ge(111)-(2x1) surface must have, in particular, two contributions. One
quasi-1D, coming from surface reconstruction, and another one, coming from
ionized donors with large localization radii. These two contributions are
superimposed on each other. The resulting STM image would be linear structure,
caused by (2x1) reconstruction, with wide spots, originating from impurities.
These spots are poorly visible because of perfect screening by $\pi$-bonded
electrons, but some influence should exist. The calculations for impurity deep
below the surface are to be done in the future and there is need to re-analyze
experimental data thinking in this direction. Also we should mention, that the
above speculations must be taken with great care, as the common sense might
often be misleading in surface physics.

The overall behavior of impurity $LDOS$ images strongly differs, depending on
the position of donor atom in the crystal cell. This is observed even for
subsurface defects. The same P donor impurity may looks as protrusion,
depression, protrusion superimposed on a depression etc. It depends both on the
spatial location of impurity and applied bias voltage.

The atomic orbitals in vicinity of surface defects are strongly hybridized.
This results in up/downward band edges "bending''. The insets of
Fig.~\ref{LDOS_maps} with split state areas zoomed in with high contrast,
illustrate this. Basically, the orbital's hybridization leads to specific
spatial shape of tunneling spectra $LDOS(x, eV)$ and, in other words, to the
appearance of local electronic density spatial
oscillations~\cite{JETP_LETT2011, JETP_LETT2012}.
Let us note, these are not charge density oscillations, because they are
observed in empty states energy range (above Fermi level). Spatial $LDOS$
oscillation on Ge(111)-(2x1) surface were the subject of~\cite{SWP} work.

The energy differences, measured with respect to the total energy of a system
of 2646 atoms with P donor atom at position 7, are shown on every pane of
Fig.~\ref{LDOS_maps}. The difference is not very large. At least, we suppose,
it does not allow to make any conclusions about the most favorable position of
donor atom. The difference is large for donor positions  8. We do not have
any explanation for huge energy gain for impurity position 8. At the same time
this energy difference applies to the huge surface slab. Due to slightly
different
atoms relaxation a few electronvolts can easily be acquired by the whole
super-cell. Also it could be, that the thickness of model slab is
not sufficient.

The last row of Fig.~\ref{LDOS_maps} will be described later on.

\begin{figure}[!h]
\leavevmode
\centering{
\includegraphics[width=80mm]{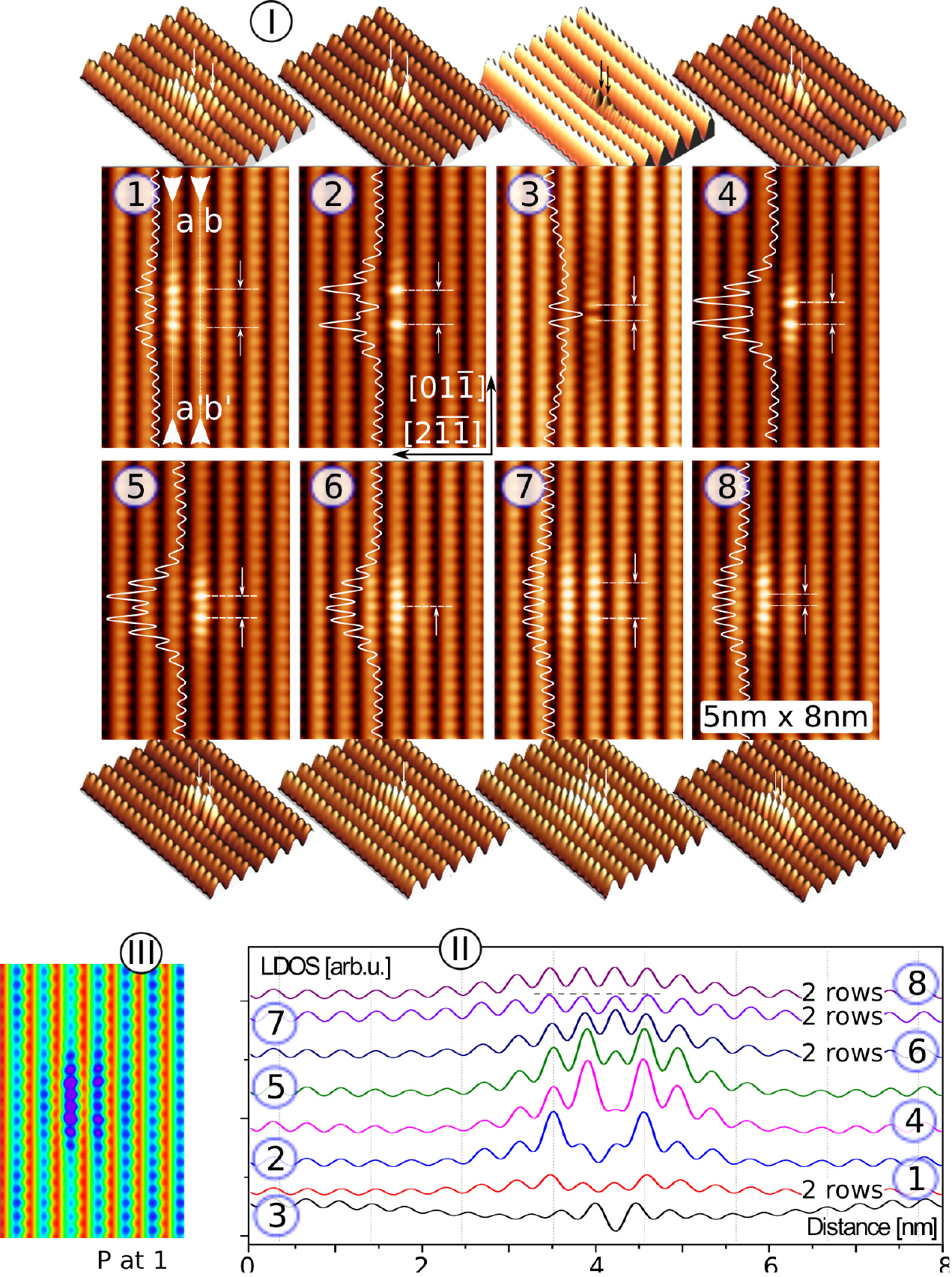}}
\caption[]{\label{LDOS_images}
{\bf (I)} $LDOS(x, y)$ maps in vicinity of P atom, located at different
positions in subsurface layers of Ge(111)-(2x1) surface. Numbers denote atoms
position. Maps are given for zero bias voltage. Profiled along b-b' line are
sketched on maps. White lines and arrows are marking the positions of maxima in
dimer row, nearest to the impurity atom. Crystallographic directions and
images size are indicated on the figure. Lines a-a' and b-b' are the same as on
Fig.~\ref{Overview}.
{\bf (II)} The profiles of $LDOS(x, y)$ maps along b-b' line on the same scale.
Note the shift of profile for position 2.
{\bf (III)} High contrast $LDOS(x, y)$ map for impurity at position 1,
illustrating the disturbance of surface electronic structure in a few
$\pi$-bonded dimer rows (see text).}
\end{figure}

The $LDOS$ (and STM) image of individual impurity is dominated by the split
state at zero (Fig.~\ref{Bands}) and low bias (see above) voltage as
illustrated by Fig.~\ref{LDOS_images}, where zero bias maps of
$LDOS(x, y)|_{eV=0}$ for different donor atom positions are presented together
with corresponding quasi-3D images. The profiles along b-b' direction
(the same for Fig.~\ref{Overview} and Fig.~\ref{LDOS_images}) are shown on the
maps with equal scales.

Three things can be immediately noticed from Fig.~\ref{LDOS_images}. Firstly,
one or two $\pi$-bonded rows are affected by donor impurity. Secondly, one or
two local maxima are present on the profile. Thirdly, the distance between
maxima can be one or two dimers along $\pi$-bonded row ([01$\rm{\overline{1}}$]
direction). This is shown in Fig.~\ref{LDOS_images} by thin lines and arrows and
is summarized in Table.~\ref{Table1}.

\begin{table}[!h]
\caption{\label{Table1}%
P donor impurity $LDOS$ image properties }
\begin{ruledtabular}
\begin{tabular}{lcccccccccc}
Atom position&  &\vline &1 &2 &3 &4 &5 &6 &7 &8 \\
\colrule
\hline
2 rows&		&\vline  &x &- &- &- &- &x &x &x \\
2 maxima&	&\vline  &x &x &x &x &x &- &x &x \\
Num. dimers&   	&\vline  &2 &2 &1 &1 &1 &- &2 &1 \\
\end{tabular}
\end{ruledtabular}
\end{table}

Thus, P donor impurity at position 1 is imaged as two row feature with two
maxima in a row and double dimer distance between maxima.

To be absolutely accurate, not two, but a few rows are affected by impurity. The
situation is the same as with spatial extent of $LDOS$ image protrusion. One
should either increase the contrast of images (see Fig.~\ref{LDOS_images}(III)),
either use profiles in analyses (Fig.~\ref{LDOS_images}(II)). Reduction to
two disturbed rows allows to classify $LDOS$ images of impurity located at
different positions.

Having this classification, we can apply it to the test case. Si(111)-(2x1) and
Ge(111)-(2x1) surfaces are similar in many senses. It is possible to perform a
simple check of our results by comparison with Si(111)-(2x1)
surface~\cite{Si:P}. In general the situation with STM imaging of individual
impurities is much more simple on Si(111)-(2x1) surface. The empty surface
states band $\pi^*$ and the valence band VB are separated by about 0.4~eV
gap~\cite{Buckling1}. Near Fermi level there are no states, available for
tunneling, but empty surface states. That is why the STM impurity images on
Si(111)-(2x1) are much easier to classify.

In accordance with Fig.~\ref{LDOS_images} and Table.~\ref{Table1} the
conclusions of authors can be immediately confirmed. In our notations:
Fig.2a~\cite{Si:P} corresponds to P in position 2, Fig.2b~\cite{Si:P} - P in
position 4, Fig.2b~\cite{Si:P} - P in position 5. The remaining unclear feature
(Fig.2d~\cite{Si:P} ) most probably is the STM image of P atom, adsorbed on the
surface. This statement is out of scope of present investigation and will be
proved in the future publications~\cite{Sav}.

Now we can come back to the lowest row of Fig.~\ref{LDOS_maps}. Images 1', 6'
and 8' correspond to cross-sections of $LDOS(x, y, eV)$ scalar field along a-a'
plane (see Fig.~\ref{Overview}), i.e. along second (along with b-b') disturbed
$\pi$-bonded row. As it can be seen from Fig.~\ref{LDOS_maps} the split state
is also present on these images (see also Fig.~\ref{LDOS_surface}). Images 6'
and 8' are  similar with the main difference being the image contrast.
$LDOS$ image for the case of P donor placed at position 7 is not shown as
$LDOS$ maps along a-a' and b-b' dimer rows are almost identical. At the same
time for other cases the difference between a-a' and b-b' maps is rather big.

The only exceptional case among $LDOS(x, eV)$ 1-8 images is the case of
impurity at position 3, when the split state goes below Fermi level. To check
if this situation can be reproduced with slightly different atomic environment,
we have performed calculations for two impurities located at different
positions in the atomic lattice. First impurity was fixed at position 6, while
the second was sequentially placed at positions from 1 to 4. The results for
P4-P6 pair is depicted in Fig.~\ref{LDOS_maps}(4''). One can see that split
state for impurity at position 4 was shifted below Fermi level by adding second
impurity to position 6. Thus we proved, that split state location below Fermi
energy can be observed for different conditions. The comprehensive analyses of
donor pairs is out of scope of current paper.

\subsection{Local tunneling spectroscopy}

The last problem we would like to discuss is the local spectroscopy $LDOS(eV)$
curves. This article was written with experimenter's needs in mind, so we will
analyze spectroscopy curves as if they were obtained by STM. There are two
approaches to measure the tunneling spectra. One is simple I(V) curve
measurement at certain surface point. It heavily relies on very high stability
of mechanical system. Basically this is a case last years. Another approach is
based on averaging of I(V) curves above some surface area. It is less sensitive
to different noise.

\begin{figure}[!h]
\leavevmode
\centering{
\includegraphics[width=80mm]{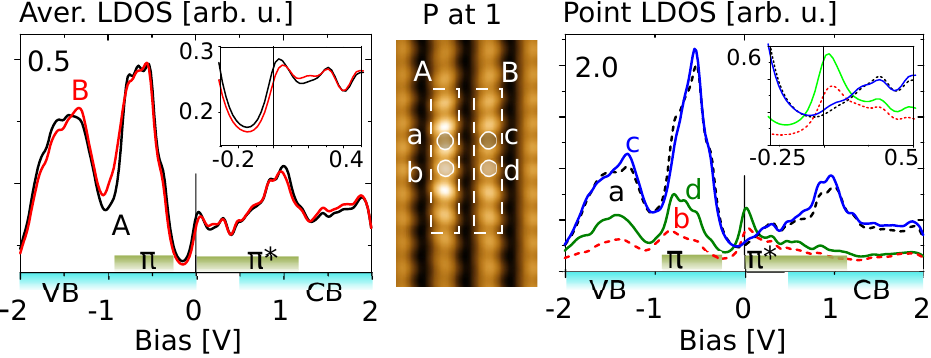}}
\caption[]{\label{Spectr} Theoretical tunneling spectra obtained by averaging
over A and B areas (left pane) and by point measurement (right pane) at a, b, c
and d points. Insets depict the zoomed in part of $LDOS(eV)$ curves around
Fermi energy. The middle part sketches the surface area above which the
spectroscopy was performed. Areas A and B are located above a-a' and b-b'
$\pi$-bonded dimer rows (see Fig.~\ref{Overview}), respectively. Points b and d
are located on top of dimers in a row, while points a and c lie in between
dimers. The band structure is shown on abscissa axis.}
\end{figure}

The problem is that two approaches can give different from the first sight
results. In Fig.~\ref{Spectr} we present model tunneling spectra for different
measurement conditions for the case of P atom located at site 1. On the left
pane there are spectra averaged over A and B areas above $\pi$-bonded rows. On
the right pane spectroscopy curves at points a, b, c and d are depicted. Points
b and d are located on top of dimers in a row. Points a and c are in between
dimers. Curves A and B are almost undistinguishable, though the heights of
protrusions along a-a' and b-b' lines (see also Fig.~\ref{Overview}) on $LDOS(x,
y)$ image strongly differ. Split state is located close to the Fermi level and
this part of spectrum is zoomed in on the inset. The difference in averaged
spectra is on a few percent scale, which apparently is not enough to draw any
reliable conclusions. Note nonzero tunneling conductivity at Fermi energy.

As to point spectroscopy, we can easily discriminate atomic size features.
Dimers at similar positions in different dimer rows give different tunneling
spectra (right pane of Fig.~\ref{Spectr}). Even relatively small difference
above elevated features along dimer rows is obvious.

Looking at the spectra obtained by different methods we can conclude, that
point spectroscopy does not give immediate impression on the band structure,
while spectroscopy with averaging does. Averaging of two point
spectroscopy curves, one on top of dimer and another in between dimers, will
give curve, similar to averaged spectroscopy curve. Also note the vertical
scales on both panes. Averaging significantly decreases the maximum value.

There are no obvious specific points on numerical tunneling conductivity  I(V)
dependence (nor on its derivative) allowing simple determination of band gap
edges (see Fig.~\ref{Bands}(III)). I.e. having perfectly defined I(V) and
knowing $LDOS$ we can {\it not} determine the band gap, although this can be
caused by very narrow DFT band gap.

Another conclusion that can drawn from local spectroscopy analyses is that it
is almost impossible to identify individual impurity on Ge(111)-(2x1)
surface relying only on the results of local spectroscopy. As we discussed
earlier, the maps of $LDOS$ should be used together with local
spectroscopy data~\cite{JETP2002, JETP_LETT2011}.

\subsection{Model's limitations}

Let us specify the strong assumptions used in present calculations. Some of them
are imposed by very big simulation super-cell.

In particular, we have performed the simulation in LDA approximation. It is
known it gives non-physically small values of band gaps. This can be slightly
improved by the GGA approximation, but real improvements can be achieved only
with computationally expensive GW many body corrections~\cite{Buckling}. At the
same time cheap "scissors`` method works quite well~\cite{Buckling2}.

There is no correction for closed STM feedback loop. The values of $LDOS$ are
calculated on the {\it plane} above the surface.

There is no STM tip density of states in our results.

In our model we can not account for the surface band bending. We simply do not
have sufficiently thick model slab. Our slab is about 15~\AA~ thick, and the
depletion layer at Ge(111)-(2x1) surface with n-type of bulk conductivity is
almost 250~\AA~ thick. The depletion layer strongly affects the picture of
tunneling for n-type doped Ge samples~\cite{JETP_LETT2005}. The same concerns
the Si(111)-(2x1) surface.

That is why our model STM images do not coincide exactly with experimental
observations, but nevertheless the correspondence is reasonable. All $LDOS(x,
eV)$ maps (except position 3) predicts the presence of protrusion on the STM
images at zero (and small) bias voltage, which indeed agree with
experiment~\cite{Sav}. We did not find any substantial difference when
explicitly adding charge to the impurity atom.

\section{Conclusions}

In conclusion, we have performed the numerical modeling of Ge(111)-(2x1) surface
electronic properties in vicinity of P donor impurity atom located near the
surface. We have found a notable increase of surface $LDOS$ around surface
dopant near the bottom of empty surface states band $\pi^*$, which we called
split state due to its limited spatial extent and energetic position inside the
band gap. We show, that despite of well established bulk donor impurity energy
level position at the very bottom of conduction band, surface donor impurity on
Ge(111)-(2x1) surface might produce energy level below Fermi energy, depending
on impurity atom local environment. It was demonstrated, that impurity, located
in subsurface atomic layers, is visible in STM experiment on Ge(111)-(2x1)
surface. The quasi-1D character of impurity image, observed in STM experiments,
is confirmed by our computer simulations with a note that a few $\pi$-bonded
dimer rows may be affected by the presence of impurity atom.

More work is needed to clarify if deep subsurface impurity atoms will be
visible for STM on Ge(111)-(2x1) surface and to investigate how the
different surface isomers will affect STM images of individual impurity.

%
%

\begin{acknowledgments}
This work has been supported by RFBR grants and computing facilities of Moscow
State University. We would also like to thank the authors of WxSM and Chimera
free software.
\end{acknowledgments}

\appendix

\bibliography{ArXiv_2013}

\end{document}